\documentclass{iau}
\usepackage{graphicx} 
\usepackage{amsmath}

\newcommand{\src}{Swift~J1822.3$-$1606}
\newcommand{\rxte}{\textit{RXTE}}
\newcommand{\xte}{\textit{RXTE}}
\newcommand{\cxo}{\textit{Chandra}}
\newcommand{\chandra}{\textit{Chandra}}
\newcommand{\rosat}{\textit{ROSAT}}
\newcommand{\swift}{\textit{Swift}}
\newcommand{\xmm}{\textit{XMM-Newton}}
\newcommand{\tempo}{{\tt{TEMPO}}}

\newcommand{\lsk}{\cite{lsk+11}}

\newcommand\arcmin{\mbox{$^\prime$}}%
\title[IAUS291.~~The new magnetar \src] 
{The new magnetar \src} 

\author[P. Scholz \etal]  
{P. Scholz,
 C.-Y. Ng, 
 M. A. Livingstone, 
 V. M. Kaspi, 
 A. Cumming,
 \and R. Archibald} 

\affiliation{Department of Physics, Rutherford Physics Building, McGill University \\
                 3600 University Street, Montreal, Quebec, H3A~2T8, Canada \\ 
                 email: {\tt pscholz@physics.mcgill.ca}} 

\pubyear{2012}
\volume{291}  
\jname{\mbox{Neutron Stars and Pulsars: Challenges and Opportunities after 80 years}}
\editors{J. van Leeuwen, ed.} 
\begin{document}

\maketitle

\begin{abstract}
On 2011 July 14, a transient X-ray source, \src, was detected by Swift BAT via its burst activities. 
It was subsequently identified as a new magnetar upon the detection of a pulse period of 8.4\,s. 
Using follow-up \xte, \swift, and \chandra\ observations, we have determined a spin-down rate of 
$\dot{P}\sim3\times10^{-13}$, 
implying a dipole magnetic field of $\sim5\times10^{13}$\,G, second lowest among known magnetars, 
although our timing solution is contaminated by timing noise. 
The post-outburst flux evolution is well modelled by surface cooling resulting from 
heat injection in the outer crust, although we cannot rule out other models. 
We measure an absorption column density similar to that of the open cluster M17 at 10\arcmin\ away, 
arguing for a comparable distance of $\sim$1.6\,kpc. 
If confirmed, this could be the nearest known magnetar.
\end{abstract}


\firstsection 
\section{Introduction}
Over the past two decades, several new classes of neutron stars have been
discovered. 
Perhaps the most exotic is that
of the magnetars, which exhibit some highly unusual properties, often
including violent outbursts and high persistent X-ray luminosities that
exceed their spin-down powers. 

To date, there are roughly two dozen magnetars and candidates
observed\footnote{See the magnetar catalog at
\tt{http://www.physics.mcgill.ca/$\sim$pulsar/magnetar/main.html}.}, with spin
periods between 2 and 12\,s, and high spin-down rates that generally suggest
dipole $B$-fields of order $10^{13}$ to $10^{15}$\,G.
\emph{Swift} has discovered several new magnetars 
in recent years via their outbursts
\cite[(e.g. G{\"o}{\u g}{\"u}{\c s} {\etal} 2010; Kargaltsev {\etal} 2012)]{gcl+10,kkp+12}. 

One of the latest additions to the list of magnetars is \src. This source was
first detected by \emph{Swift} Burst Alert Telescope (BAT) on 2011 July 14
(MJD~55756) via its bursting activities \cite[(Cummings {\etal} 2011)]{cbc+11}. 
It was soon identified as a new magnetar 
upon the detection of a pulse period $P$=8.4377\,s \cite[(G{\"o}{\u g}{\"u}{\c s} {\etal} 2011)]{gks11}. 
In \cite{lsk+11}, we
reported initial timing and spectroscopic results using
observations from \swift, \emph{Rossi X-ray Timing Explorer} (\emph{\rxte}),
and \emph{Chandra X-ray Observatory}. We found a spin-down rate of $\dot
P=2.54\times 10^{-13}$ which implies a surface dipole magnetic
field
\footnote{The surface dipolar component of the $B$-field can be estimated by
$B=3.2\times10^{19}(P\dot P)^{1/2}$\,G.} 
$B=4.7\times 10^{13}$\,G, the second
lowest $B$-field among magnetars.
Using an additional 6 months of \swift\ and \xmm\ data, \cite{rie+12} 
present a timing solution
and spectral analysis. They find a spin-down rate of $\dot P=8.3\times 10^{-14}$
which implies $B=2.7\times 10^{13}$, slightly lower than 
that found in \lsk. \cite{snl+12} present an updated timing solution and latest
flux evolution using 46 observations from \swift/XRT, 32 observations from \rxte/PCA,
and 5 observations from \chandra/ASIS spanning more than a year.
A single archival \rosat/PSPC observation is also analysed.
In these proceedings we summarize the results of \cite{snl+12}.

\section{Results}
\subsection{Timing Behaviour}
\label{sec:timing}

\begin{table}[b]
\centering
\caption{Timing solutions for \src.}
\begin{tabular}{|l|c|c|c|}
\hline
Parameter & Solution 1 & Solution 2 & Solution 3 \\
\hline
$\nu$ (s$^{-1}$)                   &  ~~0.1185154253(3)~~	& ~~0.1185154306(5)~~    & ~~0.1185154343(8)~~ \\
$\dot{\nu}$ (s$^{-2}$)             & $-9.6(3)\times10^{-16}$    & $-2.4(1)\times10^{-15}$    & $-4.3(3)\times10^{-15}$ \\
$\ddot{\nu}$ (s$^{-3}$)            & -                          & $1.12(8)\times10^{-22}$    & $4.4(6)\times10^{-22}$ \\
$\dddot{\nu}$ (s$^{-4}$)           & -                          & -                          & $-2.2(4)\times10^{-29}$ \\
$\chi^2_\nu/\nu$                   & 5.02/72                    & 1.94/71                    & 1.44/70 \\
$B$ (G)                            & $2.43(3)\times10^{13}$     & $3.84(8)\times10^{13}$     & $5.1(2)\times10^{13}$ \\
\hline
\end{tabular}
\label{ta:coherent}
\end{table}

For each \swift\ and \cxo\ observation, a pulse time-of-arrival (TOA) was 
extracted using a Maximum Likelihood (ML)
method, which yields more accurate TOAs than the
traditional cross-correlation technique \cite[(see Livingstone {\etal} 2009)]{lrc+09}. 
For the \rxte\ observations the cross-correlation method was used,
as the high number of counts make the ML method computationally expensive.

Timing solutions were then fit to the TOAs using \tempo\
We fit three solutions, one with a single frequency derivative (Solution 1), 
one with two derivatives (Solution 2) and one with three derivatives (Solution 3). 
Table \ref{ta:coherent} shows the best-fit parameters for the three solutions.
The addition of higher-order derivatives significantly improves
the fit with the solutions having a reduced $\chi^2_\nu/\nu$ of 5.02/72,
1.94/71, and 1.44/70, respectively.
The second and third frequency derivatives serve to fit out
the effects of apparent timing noise.
The best-fit solution, with three significant derivatives, 
has a $\nu$ and $\dot\nu$ which
imply a spin-inferred dipole magnetic field of $5.1(2)\times10^{13}$\,G,
the second lowest magnetic field measured for a magnetar thus far.
This $B$-field is slightly higher than the value, $2.7\times10^{13}$\,G, measured
by \cite{rie+12} as they do not measure significant second and third frequency
derivatives. For a detailed comparison of our works see \cite{snl+12}.

\subsection{Flux Evolution}

We fitted the \swift\ and \chandra\ spectra with a blackbody plus power-law model using XSPEC
and measured 1--10\,keV fluxes. 
We find a best-fit $N_H=4.53(8)\times 10^{21}\, \textrm{cm}^{-2} $ and that the spectrum softens as
the flux decays. The flux decay can be characterised by a double-exponential model with decay
timescales of $15.5\pm0.5$ and $177\pm14$ days.

We find that the observed luminosity decay is also well reproduced by models of thermal relaxation 
of the neutron-star crust following the outburst. 
We follow the evolution of the crust temperature profile by integrating the thermal diffusion equation. 
The calculation and microphysics follow \cite{bc09} who studied transiently accreting neutron stars, 
but with the effects of strong magnetic fields on the thermal conductivity included \cite[(Potekhin {\etal} 1999)]{pbhy99}.
We assume $B=6\times 10^{13}\ {\rm G}$, similar to the value inferred 
from the spin down and a 1.6 $M_\odot$, $R=11.2\ {\rm km}$ neutron star. 

We obtain good agreement with the observed light curve for times $<100$ days 
with an injection of 
$\sim 3\times 10^{42}\ {\rm ergs}$ of 
energy at low density $\sim 10^{10}\ {\rm g\ cm^{-3}}$ in the outer crust at 
the start of the outburst (Figure \ref{fig:fluxmodel}).
This conclusion 
comes from matching the observed timescale of the decay, and is not very sensitive to the choice of 
neutron-star parameters. 
We find that it is difficult to match the observed light curve at times $\gtrsim 200$ days, 
but the late time behaviour is sensitive to a number of physics inputs 
associated with the inner crust.
We will investigate the late-time behaviour in more detail in future work. 

\subsection{Distance Estimation}

\label{sec:distance}

\begin{figure}
 \begin{minipage}[t]{0.55\textwidth}%
 \centering
 \includegraphics[height=50mm]{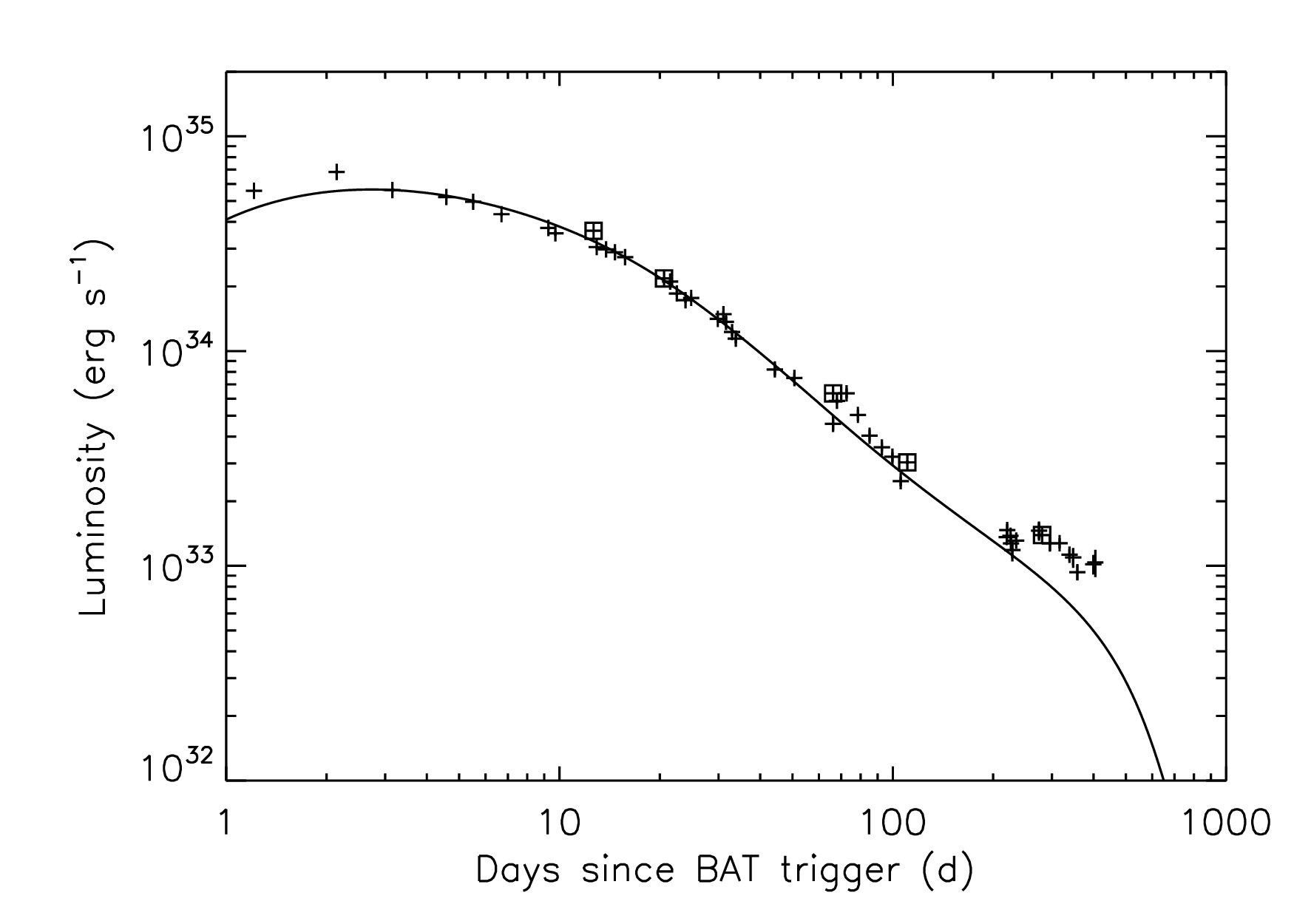}
 \caption{
  A model of the thermal relaxation of the neutron-star crust 
  that approximately reproduces the observed 1--10 keV luminosity decay assuming 
  a distance of 1.6\,kpc. 
  }
 \label{fig:fluxmodel}
\end{minipage} ~
\begin{minipage}[t]{0.42\textwidth}%
 \includegraphics[trim=5 0 0 6, clip,height=50mm]{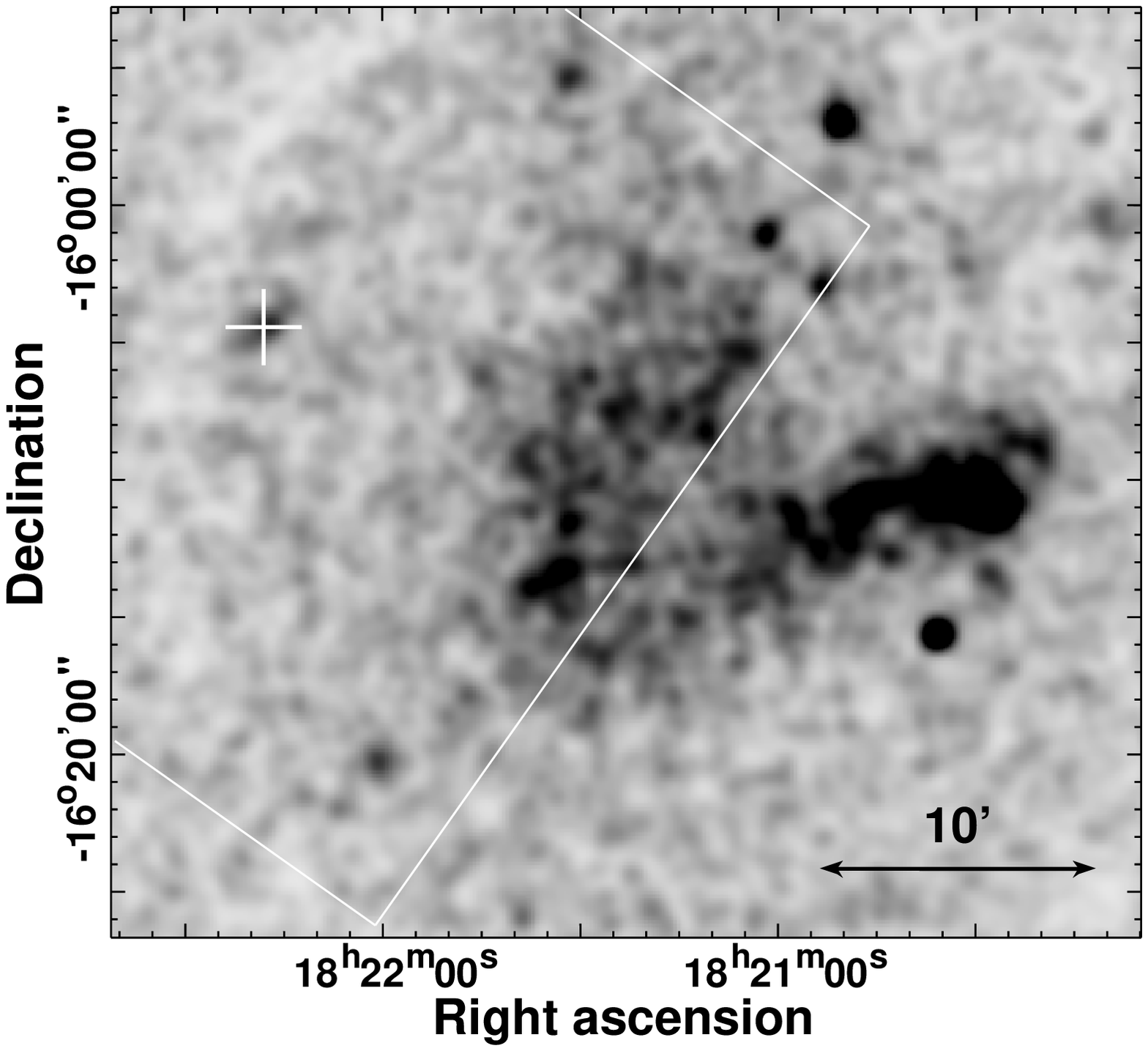}
 \caption{
  \rosat\ image of the field of \src\ in the 0.1--2.4\,keV
  range. The position of \src\ is marked by a cross.
  The large-scale diffuse emission is the Galactic H{\sc ii} region M17.} 
   \label{fig:rosat}
\end{minipage}%
\end{figure}


As shown in the \rosat\ image (Figure \ref{fig:rosat}), the Galactic H{\sc ii} region
M17 is located $\sim$20\arcmin\ southwest of \src. It has a distance of
$1.6\pm0.3$\,kpc \cite[(Neilbock {\etal} 2001)]{ncjm01} and an absorption column density
$N_{\rm H}=4\pm1\times10^{21}\,$cm\,$^{-2}$ \cite[(Townsley {\etal} 2003)]{tfm+03} which
is consistent with our best-fit value of $4.53\times10^{21}$\,cm\,$^{-2}$.
This suggests that
\src\ could have a comparable distance to that of M17.
If so, then \src\ would be
one of the closest magnetars detected thus far. 

\section{Conclusions}
We have presented the post-outburst radiative evolution and timing behavior of \linebreak\src. 
We estimate the surface dipolar
component of the $B$-field to be $\sim5 \times 10^{13}$\,G,
although this measurement is contaminated by timing noise.
By applying a crustal cooling
model to the flux decay, we found that the energy deposition likely occurred in the outer crust
at a density of $\sim10^{10}$\,g\,cm$^{-3}$.
Based on the similarity in $N_H$ to that of the H{\sc ii} region M17, we argue
for a source distance of $1.6\pm0.3$ kpc, one of the closest distances yet inferred for
a magnetar.

\end{document}